# Critical cooling rate for the glass formation of ferromagnetic $Fe_{40}Ni_{40}P_{14}B_6$ alloy

Qiang Li

School of Physics Science and Technology, Xinjiang University

Urumqi, Xinjiang, P.R. China

**Abstract**

Bulk ferromagnetic amorphous Fe-Ni-P-B alloys in rod shape were formed by a rapid solidification technique. The largest amorphous specimen prepared had a diameter of ~2.5 mm and the corresponding cooling rate for the glass formation of this alloy system in our experiment can be estimate to be around 492.4 K/s by the method of finite-difference numerical calculation. This value is on the same order of magnitude as the critical cooling rate $R_c$ of $Fe_{40}Ni_{40}P_{14}B_6$ alloy estimated by the method of constructing the continuous-cooling-transformation (CCT) curve. It is indicated that the heterophase impurities have been eliminated well in our experiment.

## I. Introduction

Since the synthesis of the first glassy alloy by Duwez and coworkers[1] by a rapid quenching method, a great number of them have been produced over the last four decades. Metallic amorphous alloys possess many excellent properties such as excellent mechanical strength, good residence corrosion ability and attractive soft magnetic behavior. Especially as soft magnets, they are characterized to be of very low coercivity $H_c$, relative large saturation magnetization $B_s$, and small magnetostriction $\lambda_s$. In addition, their high electrical resistivity and excellent mechanical strength render them the best candidate as core materials for a transformer.

The critical cooling rate $R_c$ is defined as the least quenching rate at which a melt can be quenched to its glassy state. Traditionally, $R_c$ for ferromagnetic amorphous alloy is on the order of $10^5$ K/s and the thickness of the resulting foils or ribbons is limited to less than ~50 μm[2]. The limitation in size frequently degrades the magnetic properties of the final ferromagnetic amorphous products and restricts their applications[3]. To produce bulk materials, powder consolidation techniques were considered first. However the consolidation of amorphous powders and ribbons should proceed well below the glass temperature $T_g$ of the initial amorphous powders. It has been indicated that soft magnetic amorphous alloys typically have very high strengths at temperatures below $T_g$. As a result, it is very difficult to produce full density and well-bonded bulk amorphous alloys by means of the conventional consolidation technique.

To resolve the above problem, alloy systems with low $R_c$ are searched so as to form bulk glass. By definition, 'bulk' means the dimension of a system in any direction is larger than 1 mm. Among them, the most famous ones are Pd-Ni-P, Mg-Ln-TM, Ln-Al-TM, Zr-Al-TM, Ti-Zr-TM, Ti-Zr-TM-Be, Pd-Cu-Ni-P and Pd-Ni-Fe-P[4]. The lowest critical cooling rate $R_c$ recorded is ~0.10 K/s for the $Pd_{40}Cu_{30}Ni_{10}P_{20}$ alloy and the maximum diameter of an amorphous specimen prepared, $t_{max}$, reaches a value as large as ~72 mm[5]. Unfortunately, none of the above is a ferromagnet.

Inoue successfully synthesized bulk multicomponent ferromagnetic amorphous alloys of the systems Fe-(Al,Ga)-(P,C,B)[6] and Fe-(Al,Ga)-(P,C,B,Si)[7] by a conventional copper mold casting method and $t_{max}$ reaches ~2 mm. After that, a series of bulk ferromagnetic amorphous alloys in the form Fe-TM-B[8] (TM=IV-VIII group transition metal) with high mechanical strength and good soft magnetic properties were also synthesized by his group. It is now envisaged that these bulk soft magnetic amorphous alloys may already find roles as magnetic heads, magnetic cores IC cards and so on.

Recent works on bulk amorphous ferromagnetic alloys focus mainly on searching for new alloy systems with large glass formation ability. In order to improve the glass formation ability of ferromagnetic alloy, some non-ferromagnetic glass formation elements have to be added to alloy. However, it is pity that very often the addition of non-ferromagnetic elements degrades the magnetic properties of an alloy. On the other hand, many studies have indicated that the fluxing technique is an effective method in removing heterophase impurities from a molten specimen, enhancing its glass forming ability, GFA. Earlier Kui[9] found that by a fluxing technique, molten $Pd_{40}Ni_{40}P_{20}$ alloys can be cooled to the glass state bypassing crystallization with a cooling rate of ~1 K/s. The diameter of the largest specimen prepared is ~1 cm. More recently, Nishiyama[5] reported that by a similar treatment, a $Pd_{40}Cu_{30}Ni_{10}P_{10}$ alloy can be quenched to its glass state with $R_c$ ~ 0.10K/s. $t_{max}$ is ~72mm. Without the fluxing treatment, $R_c$ and $t_{max}$ are 1.57 K/s and 40 mm, respectively.

$Fe_{40}Ni_{40}P_{14}B_6$ alloy is the first commercial product of Allied Chemical Corporation under the tradename METGLAS 2826. It has excellent soft magnetic properties and attractive mechanical properties. Consequently, it can be used as magnetic shielding, magnetic recorder head and so on[10]. Since the reduced glass transition temperature $T_{rg}$ of $Fe_{40}Ni_{40}P_{14}B_6$ is 0.57, it was considered that this alloy could only be amorphized at cooling rates larger than ~$10^6$ K/s[11] and the corresponding thickness of the glassy ribbons is ~50 μm. However, most recently, bulk amorphous $Fe_{40}Ni_{40}P_{14}B_6$ rods with a diameter as large as ~2.5 mm had been synthesized by means of a technique of combination of the fluxing technique and a novel fast quenching method in our experiment[12]. In this paper a theoretical analysis on the critical cooling rate $R_c$ for formation of amorphous $Fe_{40}Ni_{40}P_{14}B_6$ alloy will be given.

## II. Experimental

$Fe_{40}Ni_{40}P_{14}B_6$ ingots were prepared from Fe chips (99.98% pure), Ni spheres (99.95% pure), B pieces (99.0% pure) and $Ni_2P$ ingots. The $Ni_2P$ ingots were themselves prepared from $Ni_2P$ powders (98% pure). After weighing in the right proportions, they were put in a clean fused silica tube and alloying was brought about by rf induction furnace under argon Ar atmosphere. All the as-prepared ingots had a mass of ~3.0 g.

Bulk ferromagnetic Fe-Ni-P-B glasses had attempted to be prepared via three different routes. In the first route, a $Fe_{40}Ni_{40}P_{14}B_6$ melt that had been fluxed was subjected to J-quenching technique. Here 'flux' means fluxing technique, in which a molten alloy was immersed in a molten oxide (anhydrous $B_2O_3$ was employed here) at an elevated temperature for a prolonged period. The oxide or fluxing agent served to remove impurities from the molten specimen. And the details about J-quenching technique can be found elsewhere[12]. In the second route, a $Fe_{40}Ni_{40}P_{14}B_6$ melt without the fluxed treatment was subjected to J-quenching technique.

In the third route, a $Fe_{40}Ni_{40}P_{14}B_6$ melt were directly quenched in the process of fluxing. First a small portion with the mass of 60~120 mg was divided from the raw as-prepared $Fe_{40}Ni_{40}P_{14}B_6$ ingot. Then it was fluxed in a fused silica tube with the inner/outer diameter of 7/9 mm under the vacuum of ~$10^{-2}$ Torr at an elevated temperature of ~1350 K. These small specimens would be melted to form a sphere with the diameter of 1.2~1.6 mm in fluxing treatment. After a fluxing treatment for a few hours, the whole system was quenched directly in the salted ice water. The cooling rate in this route had also been estimated by the following experimental method. A K-type thermocouple which has a head with the diameter of ~1.5 mm was used to substitute for the molten specimen in the above setup and was immersed in molten $B_2O_3$ fluxing agent. The thermocouple was connected to a temperature meter which data were recorded by a PC at a sampling rate of 3 points per second. Repeating the above quenching process, a temperature profile could be recorded. The cooling rate within the temperature range of 1184 K to 808 K can be determined to be around 80 K/s from the temperature profile.

The amorphization of the as-formed rod specimens was checked by X-ray diffractometer with Cu $K_\alpha$ radiation and transmission electron microscopy (TEM). The thermal behavior of as-prepared specimens was examined at a heating rate of 0.33 K s$^{-1}$ by differential scanning calorimetry (DSC) and differential thermal analysis (DTA).

**III. Results**

$Fe_{40}Ni_{40}P_{14}B_6$ alloy rod specimens with the different diameters had been prepared by means of J-quenching technique in the first and second routes. These as-prepared alloy rods had been checked by XRD, TEM and DSC. It was pointed out that the whole or the middle part of these as-prepared $Fe_{40}Ni_{40}P_{14}B_6$ alloy rods obtained from both two kinds of ingots, one fluxed and the other not, are amorphous[12]. For the rod specimens prepared from fluxed ingots, when its diameter d ≤ 1.5 mm, the whole rod could be amorphous and the length could exceed 10 cm; when d > 1.5 mm, frequently both ends of the rods became crystalline and only the middle part with the length varying from 3 to 8 cm still remained amorphous; when d > 2.5 mm, no such rods that could be remained amorphous, could be prepared. On the other hand, for these rod specimens prepared from un-fluxed ingots, when d ≤ 1.2 mm, only the middle portion of as-prepared rod is amorphous and its lengths could stretch from 3 cm to 6 cm for our experimental arrangement described earlier; when d > 1.2 mm, no amorphous could be detected along the whole rod. The details can be found elsewhere[12]. Typical DSC thermals scan of the as-prepared amorphous rods at a heating rate of 0.33 K s$^{-1}$ is shown in Fig. 1. The glass transition $T_g$ and the kinetic crystallization temperature $T_x$ can be determined to be 659.0 K and 689.8 K respectively from this thermal scan.

Moreover the results of XRD, TEM and DSC have no significantly differences between amorphous specimens obtained from fluxed or un-fluxed ingots. And these results of amorphous rods with the different diameters have also no significantly differences.

As mentioned in the experimental part, it had been attempted to prepare amorphous $Fe_{40}Ni_{40}P_{14}B_6$ alloy via the third route. But such an attempt was failure even when the diameter of molten specimen was as small as 1.2 mm.

The experimental results are summarized in Table 1.

**IV. Theoretical analysis of the critical cooling rate of $Fe_{40}Ni_{40}P_{14}B_6$**

It has been indicated above that bulk $Fe_{40}Ni_{40}P_{14}B_6$ amorphous alloy rods can be synthesized at a very low cooling rate by means of J-quenching technique. Thus it is necessary to reinvestigate the critical cooling rate $R_c$ of $Fe_{40}Ni_{40}P_{14}B_6$ alloy. The critical cooling rate $R_c$ can be evaluated based on homogenous nucleation mechanism. The most common method is to make use of the continuous-cooling-transformation (CCT) curve developed by Uhlmann[13] for the evaluation of $R_c$ of some inorganic and metallic glasses.

When the volume fraction of crystallized material $X$ in an undercooled liquid alloy is small, $X$ can usually be described very well by the Johnson-Mehl-Avrami equation:

$$X = \frac{\pi}{3} I_s U^3 t^4 \tag{1}$$

where $I_s$ is the steady-state nucleation frequencies, $U$ is the crystal growth velocity, and t is the time taken for $X$ to appear.

The steady-state nucleation frequencies $I_s$ can be written as[14]:

$$I_s = \frac{8n^{*2/3} N_A k_B T}{\eta(T) a_0^3 V_m} Z \exp(-\frac{\Delta G_{n^*}}{k_B T}) \quad (2)$$

where $\eta(T)$ is the viscosity of molten alloy, $a_0$ is the atomic jump distance, $V_m$ is the molar volume, $k_B$ is the Boltzmann constant and $N_A$ is the Avogadro's number. To evaluate some of these parameters, the density of bulk amorphous $Fe_{40}Ni_{40}P_{14}B_6$ rod has been measured by means of immersed water method at room temperature. It is equal to $7.41 \times 10^3$ kg/m$^3$. For simplicity, the temperature dependence of the density is ignored. Now the molar mass of $Fe_{40}Ni_{40}P_{14}B_6$ alloy is $50.58 \times 10^{-3}$ kg/mol. Thus it can be deduced that $a_0 = 2.25$ Å and $V_m = 6.84 \times 10^{-6}$ m$^{-3}$. The factor $Z$ is called the Zeldovich factor and in most cases, $0.01 \leq Z \leq 0.1$. Here $Z = 0.05$ is chose. $n^*$ is the critical cluster size and is given by:

$$n^* = \frac{32 N_A \pi}{3 V_m} \frac{\sigma^3}{|\Delta G_v^3|} \quad (3)$$

where $\Delta G_v$ is the Gibbs free energy difference from the liquid phase and the solid phase per unit volume, and $\sigma$ is the interface energy between the solid phase and liquid phase per unit area. $\Delta G_{n^*}$ is the nucleation barrier for the formation of the critical nucleus and is given by:

$$\Delta G_{n^*} = \frac{16\pi}{3} \frac{\sigma^3}{\Delta G_v^2} \quad (4)$$

The crystal growth velocity U can be expressed by[15]:

$$U = \frac{f k_B T}{3 \eta(T) a_0^3} \left[1 - \exp\left(-\frac{\Delta G_v}{k_B T}\right)\right] \quad (5)$$

where $f$ is the fraction of sites on the interface where atoms may preferentially be added and removed. For metallic materials, $f$ can be expressed as $f \approx 0.2 \Delta T_r$ [13], where $\Delta T_r$ is the reduced undercooling defined as $(T_m-T)/T_m$.

The overall transformation kinetics of the amorphous alloy can be constructed if the $\Delta G_v$, $\sigma$ and $\eta$ can be determined. Since measurements of the difference between the specific heat capacities of the liquid and crystal phase are difficult to make, some approximated theoretical expressions for $\Delta G_v$ have to be employed. According to Spaepen and Thompson[16]:

$$\Delta G_v = \frac{\Delta H_f (T_m - T)}{V_m T_m} \frac{2T}{T_m + T} \tag{6}$$

where $\Delta H_f$ is the molar fusion heat and it can be determined to be 12.955 kJ/mol for $Fe_{40}Ni_{40}P_{14}B_6$ by DTA measurement.

The interfacial free energy of the solid/liquid interface, σ, according to the broken-bond model, can be written as[17]:

$$\sigma = \alpha \frac{\Delta H_f}{(N_A V_m^2)^{1/3}} \tag{7}$$

where $\alpha$ is the packing factor for the given structure. For glass formation alloys the nucleation barrier $\Delta G_{n*}$ is about 60 $k_B T$ at $\Delta T_r=0.2$. Thus $\alpha$ can assumes a value of 0.41.

In order to construct time-temperature-transformation (TTT) curve, the temperature dependence of the viscosity $\eta$ of $Fe_{40}Ni_{40}P_{14}B_6$ will be required. The viscosity $\eta$ can be expressed by the widely accepted Vogel-Fulcher equation:

$$\eta(T) = A \exp(\frac{B}{T - T_0}) \tag{8}$$

where $A$, $B$ and $T_0$ are constants. Two of them, $B$ and $T_0$ can be estimated by using Kissinger's method frequently. Chen[18, 19, 20] and other authors[21, 22] showed that the apparent activation energies $E(T)$ for glass transition and crystallization in some metallic and oxide glasses coincide with those for viscous flow. This coincidence implies that both the glass transition and crystallization in metallic glasses scale as that of viscosity. The apparent activation energies for viscous flow as a function of temperature are given by:

$$E(T) = R\frac{d(\ln(\eta))}{d(1/T)} = \frac{RBT^2}{(T-T_0)^2} \tag{9}$$

where $R$ is the idea gas constant.

The apparent activation energies $E$ for glass transition ($E_g$) and crystallization ($E_p$) can be obtained by using the Kissinger plot[23]:

$$\ln\left(\frac{T^2}{\phi}\right) = \frac{E}{RT} + \text{constant} \tag{10}$$

where $T$ is the transition temperature as a function of the heating rate $\phi$ in the thermal analysis. Fig. 2 shows the Kissinger plots for glass transition and crystallization of our rod specimens. The apparent activation energies $E$ for glass transition $E_g$ and crystallization $E_p$ can be determined to be 555.8 kJ/mol and 338.4 kJ/mol, respectively. By utilizing the above value of $E_g$ and $E_p$, and meantime taking $T_g$=659 K and $T_p$=694 K measured by means of DSC at a heating rate of 0.33 K/s, with Equ.9 $B$ and $T_0$ are found to be 1545 K and 558.9 K, respectively. By taking the assumptions of $\eta(T_g)=10^{12}$ Pa.s, the expression of $\eta(T)$ can be determined as:

$$\eta(T) = 1.98 \times 10^5 \exp(\frac{1545}{T-558.9}) \tag{11}$$

However, with Equ.11, the viscosity of the specimen at the melting point will be given an unrealistically large viscosity of $2.3\times10^6$ Pa.s. It is indicated that Equ.11 is not a plausible expression of the viscosity of the specimen.

It is pointed out by Chen[19] that a higher viscosity of ~10 Pa.s can be expected for easy glass-forming alloys at their melting points, which is associated with strong short-range order occurring in these alloys. For example, high viscosities of 6.93 and 6.15 Pa.s have been determined experimentally for the two well-known glass formation alloys, $Pd_{77.5}Si_{16.5}Cu_6$ and $Pd_{82}Si_{18}$, respectively, at their melting temperatures[24, 25]. For $Fe_{40}Ni_{40}P_{14}B_6$ alloy, its reduced glass transition temperature is 0.57 and it implies better glass formability. Thus, only for the purpose of estimation, it is plausible to take the assumption of $\eta(T_m)=5.0$ Pa.s for $Fe_{40}Ni_{40}P_{14}B_6$ alloy. Together with the assumptions that $\eta(T_g)=10^{12}$ Pa.s at the glass transition temperature $T_g$ and the equation of $E_g = \frac{RBT_g^2}{(T_g-T_0)^2}$, another expression of viscosity is obtained:

$$\eta(T) = 2.216\times10^{-5} \exp(\frac{9537}{T-410.3}) \qquad (12)$$

Based on the above efforts, a TTT diagram corresponding to a volume fraction of $X=10^{-6}$, which is considered to be a just-detectable concentration of crystals[13], can now be constructed and is shown in Fig.3. Besides, a CCT diagram can be constructed from TTT diagram by the method of Grange and Keifer[26] and is included in Fig.3. So the critical cooling rate $R_c$ is given by:

$$R_c \approx \frac{T_m-T_n}{t_n} \qquad (13)$$

where $T_m$ is the melting temperature, $T_n$ and $t_n$ are the temperature and time corresponding to the nose of the CCT curve, respectively. From the CCT diagram shown in Fig.3, the values of $T_n$, $t_n$ can be determined to be 808 K, 1.52 s, respectively, for $Fe_{40}Ni_{40}P_{14}B_6$ alloy. The corresponding critical cooling rate is 247.4 K/s.

Thereinafter the cooling rate in the J-quenching technique will be estimated. Since the cooling rate of the specimen in the J-quenching experiment is too fast to be direct measured by means of the experimental method, the cooling rate of the specimen in the J-quenching experiment will attempt to be determined by the method of numerical calculation. In the J-quenching experiment, the cooling procedure will be governed by the transit heat conduction equation. The system consisted of the molten specimen and the quartz tube can be considered as an infinitely long cylinder so that the distribution of the temperature field will be independent of $\theta$ and $z$. The boundary conditions in the J-quenching experiment are considered in the following section. At the interface between the specimen and the quartz tube, the heat flux should be continuous in the cooling process. At the interface between the outer wall of the quartz tube and the quenching agent, the heat transfer in the quenching agent should be processed by means of the convection mode. With respect to these boundary conditions, the heat conduction equation can be solved by the finite-difference method. The details and parameters settings in the calculation can be found elsewhere[27]. In our experiment, the maximum diameters of bulk amorphous $Fe_{40}Ni_{40}P_{14}B_6$ alloy rods obtained from the fluxed ingots is ~2.5 mm and the corresponding average cooling rate $R_n$ within the temperature range of $T_m$(=1184 K) to $T_n$(=808 K) is calculated to be 492.4 K/s. For un-fluxed specimen, the maximum diameter of the bulk amorphous alloy rods is 1.2 mm and the corresponding calculated average cooling rate $R_n$ is 1434.5 K/s.

**V. Discussions**

In the above theoretical analysis the critical cooling rate $R_c$ for the glass formation of $Fe_{40}Ni_{40}P_{14}B_6$ alloy system can estimated to be 247.4 K/s based on homogenous nucleation mechanism. However such the critical cooling rate $R_c$ can hardly be achieved under the laboratory conditions because heterogeneous impurities cannot be avoided entirely. In this work, for the fluxed specimen, $R_c$ can be estimated ~492.4 K/s by the method of the numerical calculation and is close to the theoretical estimated $R_c$. It is indicated that heterophase impurities in our specimen have been properly eliminated in J-quenching technique.

In addition it has been mentioned above that bulk amorphous $Fe_{40}Ni_{40}P_{14}B_6$ alloy cannot be prepared via direct quenching of the molten specimen in the process of fluxing in our experiment. Compare with J-quenching technique, obviously, there are fewer heterophase impurities but the slower cooling rate in this method. In this method the cooling rate within the temperature range of $T_m$ (1184 K) to $T_n$ (808 K) can be determined to be lower than 80 K/s via the experimental measurement. Although the molten specimen is replaced with the K-type thermocouple during measurement, it can be concluded that the cooling rate of the molten specimen should be closed to the measured value based on the following two reasons: first, the thick quartz tube wall is main factor to limit the cooling rate due to the poor thermal conductivity of the fused quartz tube in our experiment; second, K-type thermocouple is alloy and so its thermal capacity is similar to that of our specimen. This experiment fact coincides well with the result of theoretical analysis. Because the cooling rate in this method is lower than theoretical estimated $R_c$, it cannot be expected to synthesize amorphous $Fe_{40}Ni_{40}P_{14}B_6$ alloy by means of this method although the heterophase impurities has been controlled more effectively than that in J-quenching technique. Meanwhile it is also revealed that the critical cooling rate calculated in our theoretical analysis is on the appropriate order of magnitude.

By means of J-quenching technique, bulk amorphous $Fe_{40}Ni_{40}P_{14}B_6$ alloy rods with a diameter as large as ~ 1.2 mm can also be prepared from un-fluxed ingots. The corresponding $R_c$ is ~1434.5 K/s by the method of the numerical calculation and is almost one order of magnitude larger than $R_c$ for the fluxed ingots. It is pointed out again that fluxing is an effective technique to remove the impurities in the alloy system.

Finally it should also be noticed that crystallization always starts at the two ends of a rod specimen in J-quenching technique. This is different from that in the copper mold casting technique performed by A. Inoue's group in which the crystallization usually starts at the interface between the specimen and the copper mold wall[4]. It means that the impurities in the free surfaces at the two ends of the molten rod are more potential nuclei than the interface between the molten specimen and the quartz tube wall. It is pointed out that the inert quartz tube is essential in J-quenching technique although its poor thermal conductivity limits the cooling rate of the specimen.

It is generally believed that the glass formation ability of alloy can be scaled well by two empirical criterions. One is the reduced glass temperature $T_{rg}$ which is defined as the ratio of the glass transition temperature $T_g$ to the melting temperature $T_m$ and proposed by Turnbull[28]. And another one is the undercooled liquid region $\Delta T_x$ which is suggested by Inoue[29] and defined as the difference between the kinetic crystallization temperature $T_x$ and $T_g$. The $T_{rg}$ and $\Delta T_x$ of $Fe_{40}Ni_{40}P_{14}B_6$ are 0.57 and 30 K, respectively, implying moderate glass formation ability. Earlier it was generally considered that the critical cooling rate $R_c$ for the glass formation of $Fe_{40}Ni_{40}P_{14}B_6$ alloy was larger than $10^6$ K/s[11]. However, in the present study, the critical cooling rate for the glass formation of $Fe_{40}Ni_{40}P_{14}B_6$ alloy after fluxing is estimated to be on the order of $10^2$ K/s. Such a large improvement should be due to the removal of heterophase impurities by the fluxing technique and the novel J-quenching technique. If the critical cooling rate for the glass formation of alloy can really be scaled well by the above two empirical criterions, both the success of our experiment and the theoretical analysis point out that the critical cooling rate of alloy systems with the moderate glass formation ability is not so high as the past expected value. There will be considerable chances to synthesize bulk amorphous alloys even from those alloy systems with lower glass formation ability if heterogeneous impurities are effectively removed and controlled in the synthesis process. This result has important significance on design and synthesis of bulk amorphous alloys, especially bulk ferromagnetic amorphous alloys, because there are many ferromagnetic alloy systems with the excellent magnetic properties but the low glass formation ability. It is pointed out that, besides optimization of the composition of alloy system, elimination of heterogeneous impurities is another important and efficient route to prepare of bulk ferromagnetic amorphous alloys.

## VI. Conclusion

This critical cooling rate $R_c$ for the glass formation of $Fe_{40}Ni_{40}P_{14}B_6$ alloy system can estimated to be 247.4 K/s by the method of constructing the continuous-cooling-transformation (CCT) curve. It is revealed that bulk ferromagnetic amorphous $Fe_{40}Ni_{40}P_{14}B_6$ alloys in rod shape had been prepared by J-quenching technique in our experiment. The largest amorphous specimen prepared had a diameter of ~2.5 mm and the corresponding critical cooling rate $R_c$ within the temperature range of $T_m$ (1184 K) to $T_n$ (808 K) can be estimate to be around 492.4 K/s by the method of finite-difference numerical calculation. This value is on the same order of magnitude as the above theoretical calculated $R_c$ and it is indicated that the heterophase impurities have been controlled well in our experiment. The success of our experiment also points out that bulk amorphous alloys have quite chance to be produced from those alloy systems with lower glass formation ability if heterogeneous impurities are effectively removed and controlled in the synthesis process.

**Acknowledgement**

I thank Xinjiang University Doctoral Research Start-up Grants (No. BS050102 ) for financial support.

**Table 1 Summary of the experimental results**

| Preparation Method | Succ. Amorph-ization | Max. Diameter / Length (mm) | Corresponding Cooling Rate (K/s) |
|---|---|---|---|
| fluxed specimen + J-quenching technique | Y | 2.5 / 8 | 492.4 [b] |
| un-fluxed specimen + J-quenching technique | Y | 1.2 / 30 | 1434.5 [b] |
| direct quenching in the process of fluxing | N [a] | - | < 80 [c] |

(a) The minimums diameter of molten specimens used in this method is ~1.2 mm;

(b) Determined by means of finite-difference numerical calculation method;

(c) Measured by the experimental method mentioned in the experimental part.

**Figure Captions**

Fig. 1  A DSC scan of a bulk $Fe_{40}Ni_{40}P_{14}B_6$ amorphous alloy. The heating rate used is 0.33 K s$^{-1}$.

Fig. 2  The Kissinger plots for glass transition and crystallization of a bulk amorphous $Fe_{40}Ni_{40}P_{14}B_6$ alloy rod. The data points were determined by means of DSC thermal scans at the heating rates of 0.083 K/s, 0.17 K/s, 0.33 K/s, 0.50 K/s, 0.67 K/s, respectively.

Fig. 3  Theoretical calculated TTT and CCT curves by the method of Uhlmann's [13], together with the cooling curves for the specimens with the diameter of 2.5 mm and 1.2 mm, respectively, estimated by the method of finite-difference numerical calculation.

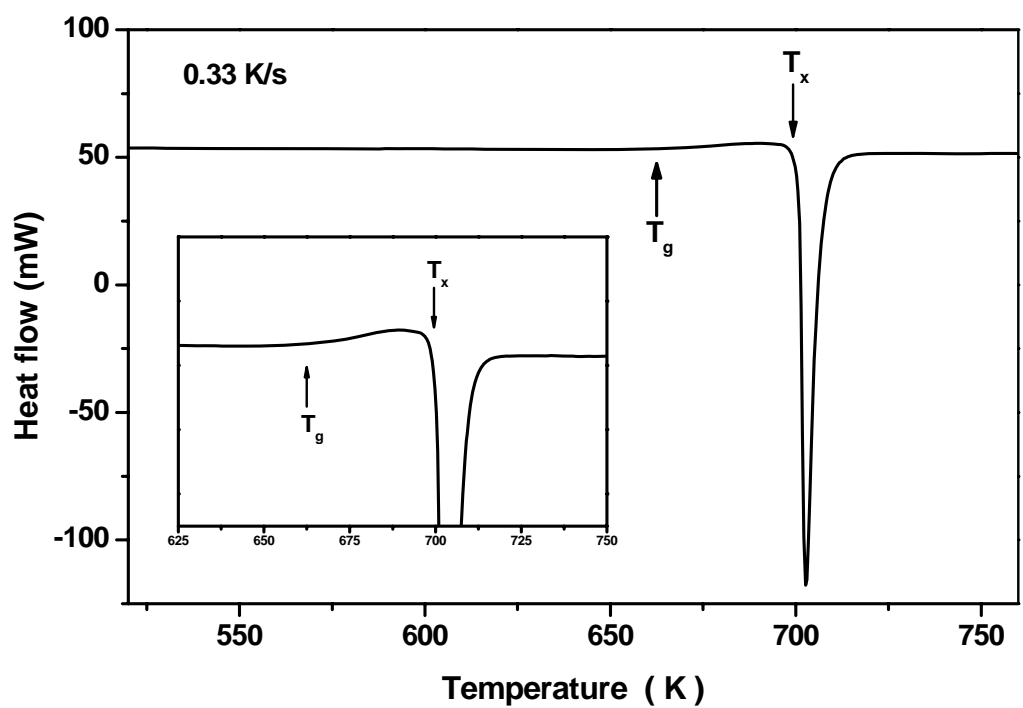

Fig.1 A DSC scan of a bulk $Fe_{40}Ni_{40}P_{14}B_6$ amorphous alloy. The heating rate used is 0.33 K s$^{-1}$.

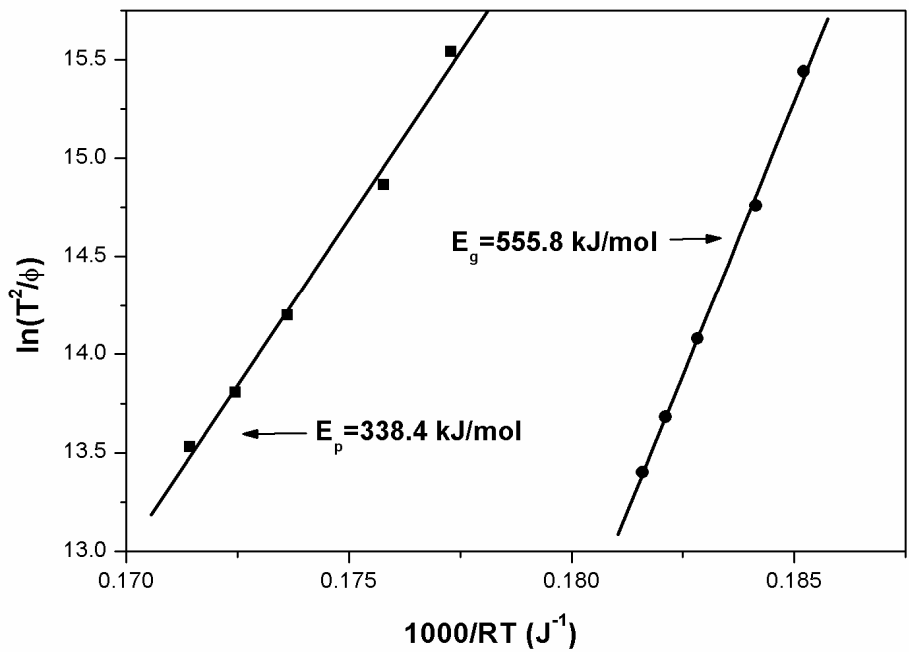

Fig.2 The Kissinger plots for glass transition and crystallization of a bulk amorphous $Fe_{40}Ni_{40}P_{14}B_6$ alloy rod. The data points were determined by means of DSC thermal scans at the heating rates of 0.083 K/s, 0.17 K/s, 0.33 K/s, 0.50 K/s, 0.67 K/s, respectively.

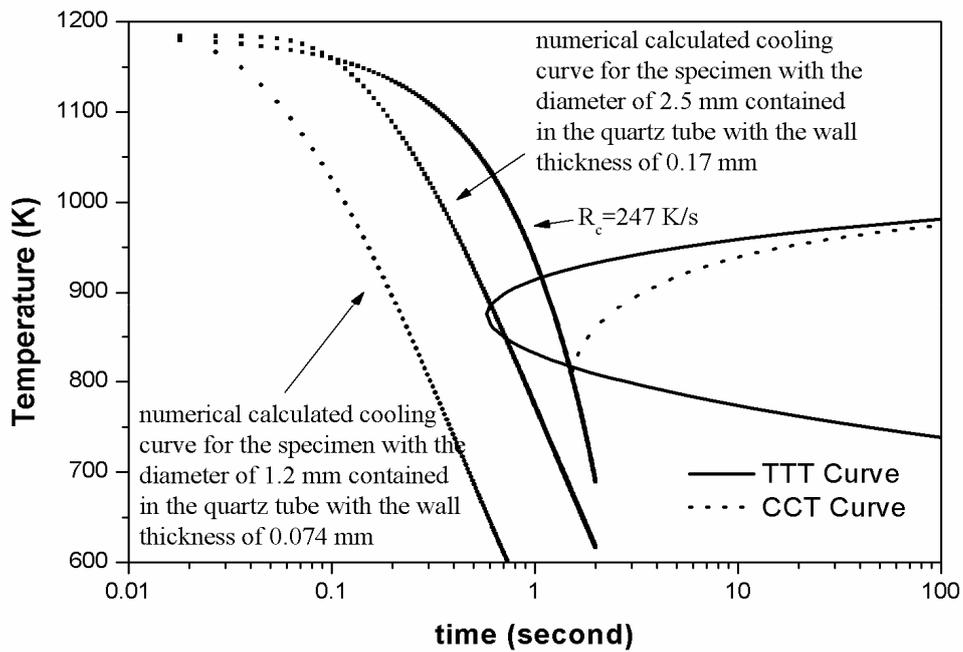

Fig.3 Theoretical calculated TTT and CCT curves by the method of Uhlmann's [13], together with the cooling curves for the specimens with the diameter of 2.5 mm and 1.2 mm, respectively, estimated by the method of finite-difference numerical calculation.